\pdfoutput=1

\documentclass[10pt,twocolumn]{article}

\usepackage[letterpaper,margin=1in]{geometry}
\usepackage{newtxtext}
\usepackage{newtxmath}

\usepackage{amsmath,amssymb}
\usepackage{mathtools}
\usepackage{enumitem}
\usepackage{booktabs,siunitx,tabularx,makecell}
\usepackage{graphicx}
\usepackage{subcaption}
\usepackage{caption}
\usepackage{placeins}
\usepackage{tikz}
\usetikzlibrary{positioning}
\usepackage[percent]{overpic}
\usepackage{algorithm}
\usepackage{algpseudocode}
\usepackage{pifont}
\usepackage{ulem}
\usepackage{etoolbox}

\usepackage[numbers,sort&compress]{natbib}

\usepackage[colorlinks=true,linkcolor=blue,citecolor=red,urlcolor=blue]{hyperref}

\usepackage[most]{tcolorbox}
\newtcolorbox{takeawaybox}{
  enhanced,
  sharp corners,
  colback=black!5,
  colframe=black!0,
  boxrule=0pt,
  frame hidden,
  left=6pt,right=6pt,top=4pt,bottom=4pt,
  borderline west={2.5pt}{0pt}{black}
}

\newcommand{\papershortname}{FOLD}

\providecommand{\hl}[1]{#1}

\begin{document}

\title{FOLD: Fuzzy Online Deduplication for Very Large Evolving Datasets via Approximate Nearest Neighbor Search}
\author{
Nelson Bore$^{1}$\thanks{Corresponding email: (\texttt{nelson.bore@mail.mcgill.ca}).}
\and
Pritish Mishra$^{2}$
\and
Constantin Adam$^{3}$
\and
Eyal de Lara$^{2}$
\and
Oana Balmau$^{1}$
\\[0.4em]
\small $^{1}$McGill University \quad
$^{2}$University of Toronto \quad
$^{3}$IBM Research
\\
}

\date{}
\maketitle

\begin{abstract}

Fuzzy deduplication is key to constructing large language model training corpora. However, classic Locality-Sensitive Hashing (LSH) pipelines scale poorly as corpora grow and are ill-suited to continuous ingestion. The main issue is that each new document batch must be checked
against the admitted corpus before insertion. As the corpus
grows, the LSH buckets grow: each query can hit several
large buckets and must scan the returned candidates. To solve this problem,  we present \ifdefstring{\papershortname}{FOLD} {\papershortname{} (Fuzzy Online Deduplication)} {\papershortname{} (Retrieval-Augmented Deduplication)}, an online fuzzy deduplication system that delivers both high recall and throughput for evolving
datasets. \papershortname{} maintains an incrementally updated HNSW index over admitted documents, retrieving a small, high-quality candidate neighborhood for each incoming document instead of repeatedly re-scanning the accumulated corpus. \papershortname{} is the first online fuzzy deduplication system to use HNSW, leading to stable throughput as datasets grow. However, it is not easy to maintain high recall when using HNSW-style indexes. The core issue is the distance metric between graph nodes. Jaccard similarity, the metric used for fuzzy deduplication, yields low recall when applied out-of-the-box with an HNSW index. It leads to distance score crowding, making graph traversal unreliable within a bounded number of steps. \papershortname{} addresses this with a bitmap representation that provides a more discriminative, Jaccard-aligned signal during HNSW search. Across four LLM-scale datasets (LM1B, C4, RealNews, and Common Crawl),
RAD preserves the scaling trajectory needed for online fuzzy deduplication:
at 30M documents, it maintains 0.94--0.97 recall relative to state-of-the-art LSH solutions, and delivers up to an 8× throughput increase.

\end{abstract}

\vspace{2pt}

\section{Introduction}
\label{introduction}
In large language model training, deduplicating the training data improves model quality and training efficiency. Fuzzy deduplication~\cite{666900, broder2000identifying} is the dominant technique used to clean text-based corpora. Unlike exact deduplication, which removes identical documents (or fragments of documents), fuzzy deduplication targets \textit{syntactic near-duplicates}. These documents share substantial text but differ due to edits, formatting changes, or copied passages. Prior work shows that fuzzy deduplication has a beneficial effect on language model training, by removing redundant examples, reducing memorization, improving generalization, and lowering training cost~\cite{hernandez2022scaling,DBLP:conf/iclr/CarliniIJLTZ23,lee-etal-2022-deduplicating,10.5555/3666122.3668470}. 

Despite its benefits, fuzzy deduplication is difficult to scale, in particular for large and continuously evolving corpora. The main challenge is posed by two factors: first, selecting a representative set of potential near-duplicate documents from terabyte-scale, increasing datasets (i.e., used by continuous model training~\cite{biesialska2020continual, DBLP:journals/corr/abs-2504-02107}), and second, efficiently computing the similarity between these documents. This paper tackles the problem of fuzzy deduplication scalability for text-based datasets, leveraging on the novel insight that graph-based approximate nearest-neighbor (ANN) search is a natural match for \textit{quickly} comparing against promising near-duplicates, at scale.

On a high-level, fuzzy deduplication represents each document as a set of $n$-grams extracted from the text (typically $n=3$--5 words~\cite{666900}) and computes a similarity score between these sets. 
The dominating similarity metric in fuzzy deduplication algorithms is Jaccard similarity~\cite{chi2017hashing}, defined as the cardinality of the set intersection divided by the cardinality of the set union. 
Because this representation is based on token and $n$-gram overlap rather than exact matches, fuzzy deduplication captures approximate textual resemblance\footnote{Note that this is a different problem from semantic equivalence, which we discuss in Section~\ref{rw}.}.
However, computing Jaccard similarity over all document pairs in large datasets is expensive at scale~\cite{broder1997syntactic}. To alleviate this issue, current fuzzy deduplication pipelines use n-gram hashing techniques~\cite{broder1997resemblance}, as well as coarse grouping of documents into clusters (also called bands, or buckets) so that only documents in the same cluster are verified~\cite{indyk1998approximate,lee-etal-2022-deduplicating,ibmdpk}.

Many pipelines provide support for fuzzy deduplication, such as Data Prep Kit (IBM)~\cite{ibmdpk}, Data Trove (Hugging Face)~\cite{datatrove}, Data-Juicer and Model Scope (Alibaba)~\cite{datajuicer,modelscope}, Red Pajama (Together AI)~\cite{redpajama}, NeMo-Curator (NVIDIA)~\cite{nvidiadatacuration}, and Milvus~\cite{milvus}. All these systems except from Milvus follow variations on the classic fuzzy deduplication pipeline described in Section~\ref{fuzzy_deduplication_background}. Milvus is a vector database system which recently provided support for fuzzy deduplication using a custom flat vector index~\cite{milvus}. All the above solutions face several important limitations. We provide an in-depth analysis and show that, as datasets scale, there is a fundamental tradeoff between high, stable throughput and high recall, illustrating this issue with IBM DPK and Milvus. As the dataset grows, document candidate buckets shift, triggering the need for re-computing the similarity between n-grams (or n-gram hash signatures) with every incoming document. Indexed systems such as Milvus, partially mitigate this issue as they do not have to recompute the buckets. However, they do not eliminate the recall--throughput tension. As we show in Section~\ref{limitations_of_fuzzy_deduplication_frameworks}, Milvus' recall directly depends on the candidate set retrieved to compute near-duplicates. A larger candidate set yields higher recall, but in turn reduces throughput, as the search is done on a flat index.

In this paper, we propose 
\ifdefstring{\papershortname}{FOLD}
{Fuzzy Online Deduplication (\papershortname{})}
{Retrieval-Augmented Deduplication (\papershortname{})},
an efficient fuzzy deduplication system for large-scale evolving datasets. At its core, \papershortname{} relies on a graph-indexed vector database to maintain the similarity relationships between the different documents in the dataset. The deduplication is done on the fly. 
When a new document is ingested, \papershortname{} uses ANN search to retrieve a small high-affinity neighborhood checks distances between documents only within that neighborhood, rather than comparing against the full corpus or rebuilding global buckets. 
Documents considered sufficiently different are then inserted into the index, allowing future batches to be checked against the accumulated corpus.

To make this approach scalable with large datasets, \papershortname{} relies on two novel techniques. First, \papershortname{} maintains a hierarchical navigable small world (HNSW) index, which is structurally different from prior fuzzy deduplication solutions--including Milvus, which uses a vector database with a flat index. Intuitively, the  search in an HNSW graph allows for millisecond-scale retrieval of candidate duplicates, as the neighbors are directly obtained from the graph, without additional verification. However, simply substituting a flat index with a graph-based index does not have a high recall out-of-the-box (as we show in Section~\ref{limitations_of_fuzzy_deduplication_frameworks}). Simply using the Jaccard similarity score in a graph-based index can lead to score ties as the graph is built, which hinders graph exploration and thus leads to low recall (as we explain in detail in Section~\ref{system_design}). To address this problem, we propose a novel bitmap-based document signature which both breaks ties when computing Jaccard similarity, and makes the distance computation amenable to parallelization (e.g., using SIMD). Together, \papershortname{}'s HNSW index combined with its novel bitmap document representation makes fuzzy deduplication scalable, while maintaining an almost perfect recall.

We evaluate \papershortname{} against IBM DPK~\cite{ibmdpk},
Milvus~\cite{milvus}, and the FAISS vector-search library using an
out-of-the-box HNSW index~\cite{hnsw_faiss}. Across four real-world
datasets---LM1B~\cite{chelba2013one}, C4~\cite{10.5555/3455716.3455856},
RealNews~\cite{10.5555/3454287.3455099}, and a Common Crawl
snapshot~\cite{commoncrawl2024-30}---\papershortname{} sustains high
DPK-relative recall with stable and high end-to-end throughput as the corpus grows.\hl{The key evaluation result is the scaling trajectory:}
\papershortname{} \hl{remains in the high-throughput, high-recall regime as the corpus grows, while Milvus loses throughput under growing candidate and maintenance costs and FAISS (Jaccard) preserves speed only with lower recall.}

In summary, this paper makes the following contributions:
\begin{enumerate}[leftmargin=*, nosep, noitemsep]
\item We perform an in-depth analysis to identify challenges of fuzzy deduplication for large, continuously evolving datasets and show that existing approaches struggle to maintain either a stable throughput, or a high accuracy, or both (Section~\ref{limitations_of_fuzzy_deduplication_frameworks}).
\item Based on the insights in Section~\ref{limitations_of_fuzzy_deduplication_frameworks}, we design and implement \papershortname{}, a fuzzy deduplication system for continuously evolving large-scale datasets. \papershortname{} is the first system to use an HNSW index over admitted documents, avoiding repeated global bucket construction or corpus candidate generation as new documents arrive. We then introduce a novel bitmap-based data representation with SIMD acceleration and cached statistics to keep document verification fast and accurate (Sections~\ref{system_design} and ~\ref{implementation}).
\item We empirically show that \papershortname{} sustains high recall and stable throughput at scale, with targeted break-down experiments isolating the sources of speedup (Section~\ref{experimental_evaluation}). \papershortname{} will be open sourced upon paper publication.
\end{enumerate}

\section{Background}
\label{background_and_motivation}
First, we introduce the basics of fuzzy deduplication. Second, we review vector database indexing, focusing on graph-based approaches. Finally, we review existing fuzzy deduplication frameworks used in our study in Section~\ref{limitations_of_fuzzy_deduplication_frameworks}.
\subsection{Fuzzy Deduplication Overview} 
\label{fuzzy_deduplication_background}
\begin{figure}
     \centering
    \includegraphics[width=\linewidth]{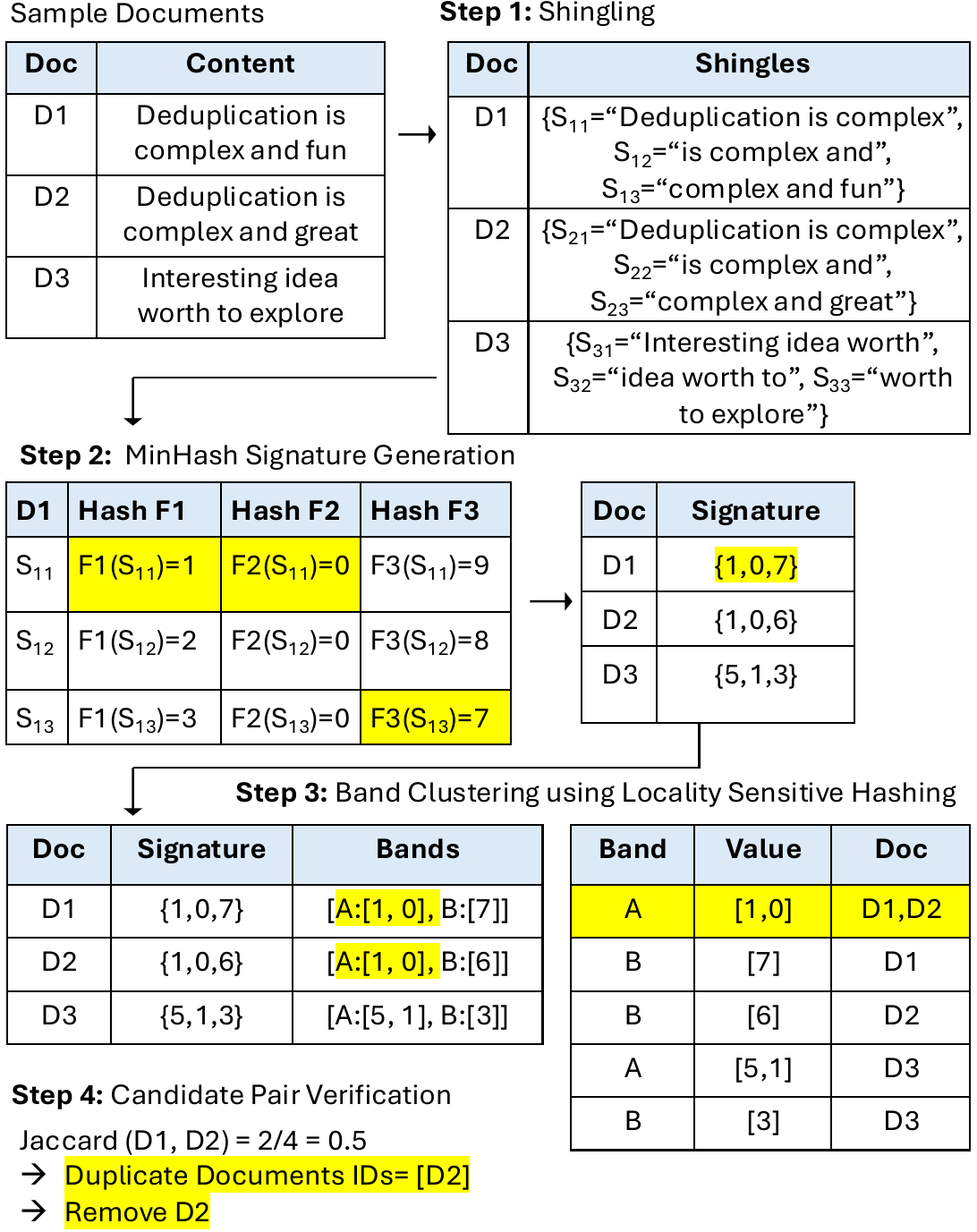}
     \caption{Steps involved in state-of-the-art Fuzzy Deduplication frameworks using MinHash and LSH.}
     \label{fig:fuzzy_dedup_flow}
\end{figure}

Figure~\ref{fig:fuzzy_dedup_flow} shows a simple example of a state-of-the-art fuzzy deduplication workflow for three documents $D_1$, $D_2$, $D_3$. The process consists of four steps:

\noindent \textbf{1. Shingling:} Each document is divided into overlapping
$n$-grams (shingles). In our example, documents are split into
3-grams, where each is built by shifting right by one word.

\noindent \textbf{2. MinHash Signature Generation:} A MinHash signature~\cite{666900} is generated per document. For each shingle $S_{ij}$ in document $D_i$, fuzzy deduplication applies a collection of hash functions. Our example uses 3 hash functions $F1$, $F2$, $F3$. In practice and in our implementation, 112 hash functions are used~\cite{ibmdpk}. For each hash function, the lowest hash value is selected across all shingles (i.e., $min(F(S_{ij})$, where $j \in 1, 2, 3$). In Figure~\ref{fig:fuzzy_dedup_flow} we detail the MinHash Signature calculation for document $D_1$. The chosen values for each hash function are $F1(S_{11})$, $F2(S_{11})$, and $F3(S_{13})$, highlighted in yellow.  
   The signature size only depends on the number of hash functions and their output size.

\noindent \textbf{3. Locality-Sensitive Hashing (LSH):} To reduce the search space, similar documents are then grouped into buckets by partitioning each MinHash signature into multiple non-overlapping bands, each containing a subset of the hash values. In our example, Band A contains the first two hash values and Band B contains the third value. Documents sharing an identical band are placed in the same bucket and non-empty buckets are forwarded to Step 4. The number of comparisons in our example is reduced from 3 to 1, with only the pair $D1$, $D2$ passed to the next step.

\noindent \textbf{4. Pair Verification:} Within each bucket, documents are compared via the Jaccard similarity~\cite{chi2017hashing} of the MinHash signatures to identify near-duplicates. Finally, a predefined threshold is used to determine whether the similarity between two documents is high enough for them to be considered near-duplicates. We use a threshold of 0.5 in our example, and $D2$ is identified as a duplicate.

As datasets grow, each incoming document must be checked against the accumulated corpus, so candidate generation and Jaccard verification become increasingly expensive. This is the key reason why classic approaches such as DPK struggle with large and evolving datasets. Vector databases, described below, offer a compelling way to support efficient candidate retrieval under continuous insertion.

\subsection{Vector Databases}
\label{vector_databases}

Vector databases store item embeddings and support similarity search at scale ~\cite{vector_survey}.
At the core of the vector database lies its main index structure, which can be implemented using different types of data structures. Each of these data structures provides a trade-off between read and insertion performance. The relevant issue for online fuzzy deduplication is how the index is able to retrieve similar documents to the newly ingested documents, as the corpus evolves, aiming for a recall higher than 0.9~\cite{DBLP:conf/iclr/CarliniIJLTZ23}. 

Flat indexes, such as FAISS Index Flat~\cite{hnsw_faiss} and Milvus FLAT~\cite{milvus_indexes}, compare a query against every stored item, giving exact results but requiring $O(N)$ work per query. Partitioned indexes, such as IVF-style indexes~\cite{faiss_git_docs,milvus_indexes}, reduce this cost by assigning items to clusters or buckets and probing only selected clusters whose centroids are closest to the incoming query. 
For large evolving corpora, these indexes are forced to either scan more data (reducing throughput), and periodically rebuild the clusters to preserve recall.

Graph-based indexes such as the
Hierarchical Navigable Small World (HNSW), build a layered proximity graph for low-latency approximate nearest-neighbor search and support online inserts~\cite{DBLP:journals/corr/MalkovY16}. 
Search greedily descends from sparse upper layers toward dense layers and then performs a bounded search in the bottom layer. Insertions connect each new item to its closest neighbors in the graph. 
 HNSW graphs rely on three key parameters: (1) \textit{M} controls the maximum number of neighbors per node, determining graph density, memory overhead, and recall; (2) \textit{efConstruction} controls the number of candidates explored during index construction, trading build time for index quality; (3) \textit{efSearch} controls the number of candidates explored during query processing, trading latency for recall.
HNSW is a natural fit for continuously evolving datasets, where each admitted document should immediately become searchable by future batches.
However, as we show in Section~\ref{limitations_of_fuzzy_deduplication_frameworks}, using HNSW out-of-the-box with Jaccard distance as the similarity metric (as required by fuzzy deduplication) does not have sufficient accuracy.

\setlength{\abovecaptionskip}{0pt}   
\setlength{\belowcaptionskip}{-10pt} 

\subsection{Existing Fuzzy Deduplication Frameworks}
\label{existing_fuzzy_deduplication_frameworks}

There are three flavors of systems used to perform fuzzy deduplication. Most existing fuzzy deduplication systems generally follow the MinHash-LSH workflow described in Section~\ref{fuzzy_deduplication_background}. In the rest of the paper, we use IBM DPK~\cite{ibmdpk} as an exponent of these frameworks. 
Second, classic set-similarity joins use frequency-ordered prefix filters to generate candidates before Jaccard verification~\cite{xiao2008efficient,xiao2009topk,xiao2011efficient,vernica2010efficient}. 
Finally, Milvus~\cite{milvus} introduced a fuzzy deduplication index,
\texttt{MINHASH\_LSH}, that implements LSH-style candidate grouping inside
the vector database. Unlike \papershortname{}, Milvus does not use
graph-based ANN search for deduplication. Instead, it retrieves candidates
from shared LSH bands, uses a Bloom filter to accelerate bucket-membership
checks~\cite{bloom_filter}, and verifies candidates using Jaccard distance.
This design moves fuzzy deduplication into an indexed retrieval system, but
it still relies on flat bucketed candidate retrieval: a small candidate
budget can miss near-duplicates outside the searched buckets, while a larger
budget increases verification work without reaching the desired recall target. We show the limitations of these
solutions in the next section.

\section{The Limitations of Current Fuzzy Deduplication Frameworks}
\label{limitations_of_fuzzy_deduplication_frameworks}

In this section, we show that existing fuzzy deduplication frameworks cannot maintain high throughput \textit{and} recall under large, evolving datasets.  We first analyze the approaches described in Section~\ref{fuzzy_deduplication_background}, and other non-graph indexed approaches. We then explore the potential of HNSW graph-based approaches for fuzzy deduplication. We show that both their recall and scalability depend
critically on the choice of the distance metric between graph nodes, as this is the core mechanism used to traverse the graph.

\noindent\textbf{Hardware.} All experiments ran on a Google Cloud \texttt{c3d highmem} VM (AMD Genoa, x86\_64) with 32 CPU cores and 480\,GB of memory.

\noindent\textbf{Baselines.} We evaluate five baselines grouped into two families:
"flat"-indexing systems (DPK, Prefix-Filter, Milvus), and graph-based systems (FAISS with HNSW indexing and two distance metrics).

\begin{itemize}[leftmargin=*, nosep, noitemsep]

\item \textbf{DPK}~\cite{ibmdpk} is an exponent of the LSH fuzzy-deduplication workflow described in Section~\ref{fuzzy_deduplication_background}. \footnote{For fairness, we SIMD-parallelize IBM DPK's band processing/candidate-set intersection, preserving candidate generation and Jaccard verification.}

\item \textbf{Milvus~\cite{milvus}} uses its custom \texttt{MINHASH\_LSH} flat index.

\item \textbf{Prefix-Filter} is our implementation of prefix-filtering set-similarity joins~\cite{xiao2008efficient,xiao2011efficient,vernica2010efficient}:
documents are 5-word shingle-hash token sets, candidates are retrieved using
rare-token prefixes, and matches are verified with Jaccard similarity.

    \item \textbf{FAISS (Hamming)} is the out-of-the-box implementation of the HNSW index from the FAISS library, using Hamming distance as the similarity metric between vertices.
    \item \textbf{FAISS (Jaccard)} is our modification of the baseline above, where we implemented the Jaccard similarity metric and use it instead of the Hamming distance. We include this baseline to measure the effectiveness of the Jaccard similarity metric used directly inside a graph-based system. 
\end{itemize}

\begin{table}[t]
\centering

\caption{Runtime and recall on a 3M-document Common Crawl snapshot, using brute-force MinHash comparison at $J \ge 0.7$ as ground truth. Brute-force requires 5 days even for a small dataset, we select DPK (the highest-recall baseline) as the practical recall reference for larger datasets. Additional workloads and larger datasets are presented in Section~\ref{experimental_evaluation}.}

\label{tab:ccmain_3m_groundtruth}
\scriptsize
\renewcommand{\arraystretch}{0.7}
\resizebox{\columnwidth}{!}{%
\begin{tabular}{lccccccc}
\toprule
  &
\makecell{\textbf{Brute}\\\textbf{Force}} &
\makecell{\textbf{DPK}} &
\makecell{Prefix\\Filter} &
\makecell{FAISS\\Jaccard} &
\makecell{FAISS\\Hamming} &
\makecell{Milvus\\topK=4} &
\makecell{Milvus\\topK=160} \\
\midrule
Time   & \textbf{5 days} & \textbf{2 hrs} & 9 hrs & 2.33 hrs & 0.66 hrs & 1.57 hrs & 3.04 hrs \\
Recall & \textbf{1.00} & \textbf{0.92} & 0.82 & 0.51 & 0.61 & 0.67 & 0.76 \\
\bottomrule
\end{tabular}%
}
\end{table}

\noindent\textbf{Experimental setting.}
Consider the input document batches $D_1,D_2,\ldots$, an existing clean corpus $U$ and a threshold $\tau$. Each incoming document $d \in D_i$  is discarded if some previously admitted document $u \in U$ satisfies $J(d,u)\ge\tau$, where $J$ is the Jaccard similarity. Otherwise, $d$ is admitted and inserted into the corpus. We evaluate recall as the fraction of near-duplicates detected and throughput as input documents processed per second.
We evaluate recall against the recall of IBM DPK. This is done because computing the exact ground truth (i.e., brute-forcing the pairwise comparison of all documents for large datasets) is prohibitively time consuming. 

To validate IBM DPK as a good-enough ground truth, we perform a brute-force search over 3M-document subsets of our evaluated datasets.
Table~\ref{tab:ccmain_3m_groundtruth} reports the recall and runtime of each  baseline against the brute-force approach at $J \ge 0.7$ for Common Crawl Snapshot ~\cite{commoncrawl}. The dataset has a ~40\% fuzzy duplicate proportion, as shown in Table~\ref{tab:dataset_text_stats} in Section~\ref{experimental_evaluation}. 
DPK achieves the highest recall among scalable baselines (0.92) while reducing runtime from 5 days to 2 hours. Similar results were obtained for the other datasets we consider, but are omitted for brevity. Therefore, in the rest of the paper, recall is measured as the fraction of DPK-detected fuzzy duplicates that are also detected by each method.

\textbf{Experiment runtime.} We perform continuous ingestion of documents in 500K-document cycles. We maintain on-disk base corpus of unique documents seen thus far and an in-memory index containing document signatures. We begin with an empty corpus. Each cycle follows two-step process. First, we perform a bulk ingest phase that deduplicates the batch of documents by filtering against the current base  corpus, as well as within the incoming 400K document batch.  Only unique documents are appended to the corpus. Second,  after the 400K-document bulk-ingest phase, a streaming-evaluation phase processes the remaining 100K documents and measures retrieval accuracy and throughput. Unless otherwise specified, indexed baselines retrieve the top 4 nearest neighbors. Throughput (documents/sec) is measured as 100K divided by the wall-clock time for the evaluation phase. Before advancing to the next cycle, each baseline adds the documents it classifies as unique to its corpus and index; this mirrors streaming deployment but can favor baselines that miss duplicates, since those missed duplicates remain in the evolving state. We first discuss flat-indexing systems in Section~\ref{sec:fuzzy-lsh}, then turn to graph-based systems in Section~\ref{graph_base_comparisson}.

\begin{figure}[t]
     \centering
         \includegraphics[width=\linewidth]{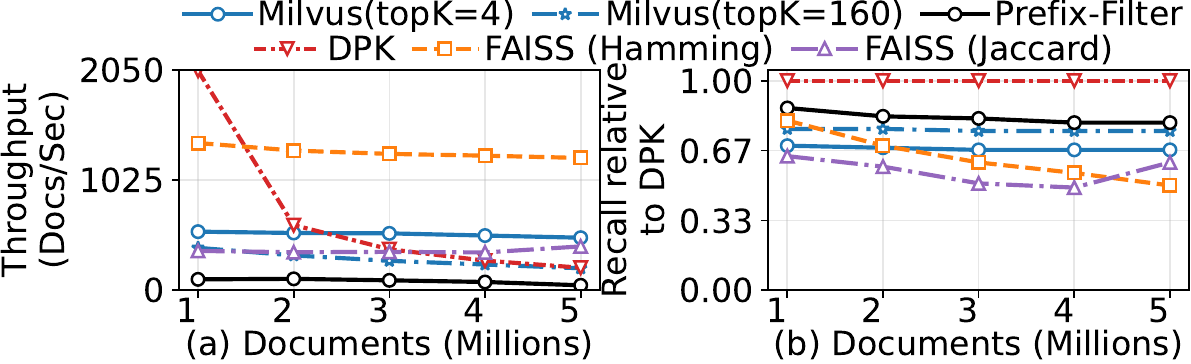}
\caption{\textbf{Throughput (left) and recall (right) for Common Crawl, as the corpus grows to 5M documents}. None of the baselines manage to maintain \textit{both} high recall and high throughput as the dataset grows.}
     \label{fig:throughput_recall}
\end{figure}

\subsection{The limitations of flat indexing methods}
\label{sec:fuzzy-lsh}

Figure~\ref{fig:throughput_recall} shows the throughput and recall (higher is better) for a 5M-document subset of Common Crawl.
First, we note that DPK's throughput drops sharply as the corpus grows: though the SIMD parallelism optimization helps when the dataset is small, each new batch must still be checked against an increasingly large corpus. 

Second, Prefix-Filter is even slower, starting at 101 documents/sec and dropping to 46 documents/sec, while its recall decreases from 0.87 to 0.80. Evolving token frequencies and growing candidate sets increase the cost of prefix-based retrieval and final Jaccard verification.

Finally, Milvus provides the most stable throughput among the flat-indexed baselines, but has low recall. Even increasing the candidate neighbors set to topK=160 improves recall to only ~70\%, while reducing throughput by ~50\%.

\begin{takeawaybox}
\textbf{Takeaway 1.} We conclude that existing approaches do not simultaneously achieve both scalability and recall under continuous ingestion: DPK-style pipelines preserve recall but slow down significantly, Prefix-Filter is slow and loses recall, and Milvus trades throughput for only a modest increase in recall.
\end{takeawaybox}

\subsection{The limitations of out-of-the-box graph indexing}
\label{graph_base_comparisson}
Vector databases using graph indexes such as HNSW provide a compelling alternative to the techniques seen above, as the  structure of the graph allows for more efficient inserts of new documents, without having to compare against increasingly growing buckets. Unfortunately, we show that such systems cannot efficiently be run out of the box, even though they have high potential.

First, we evaluate the default FAISS HNSW implementation~\cite{hnsw_faiss}, which uses the Hamming distance to compute the similarity between graph nodes. While this solution is promising for throughput, outperforming the top flat-indexed baseline by 3x, its accuracy is not satisfactory. For the largest evaluated corpus, FAISS' (Hamming) recall drops to half that of DPK, i.e., similar to a random decision of whether a document is marked as a duplicate or not.
This is unacceptable for deduplication because every missed near-duplicate is admitted into the corpus and can be repeatedly used during training, reducing the quality of the dataset.

The low recall stems from the distance metric used by HNSW out-of-the-box to assess similarity between vertices representing document signatures. By default, FAISS uses the Hamming distance. For binary signatures, the Hamming distance ($d_H$) is the number of bit positions at which two signatures differ (i.e., the number of bit flips). Many retrieval systems therefore rank candidates by minimizing $d_H$. A tempting (but invalid) proxy is to simply use Hamming similarity to cut-off incoming nodes. For example, one might use $1-d_H/B \ge 0.7$ as a proxy for a Jaccard $J \ge 0.7$\ cutoff, where $B$ is the total number of bits in the packed signature and $d_H$ is the number of differing bits. However, the meaning of the two metrics is not the same.

While Jaccard similarity depends on the fraction of \textit{identical} hash values, the Hamming distance counts bit differences \textit{within} each hash value. As a result, two signatures can share no identical MinHash values, and therefore have zero Jaccard agreement, while still appearing close under normalized Hamming similarity. This makes Hamming distance an unstable proxy for fuzzy deduplication. We provide a concrete example illustrating this mismatch in Appendix~\ref{appendix:hamming_minhash_example}.

We take our exploration a step further by adding an off-the-shelf implementation of Jaccard similarity to FAISS' HNSW index. Surprisingly, FAISS (Jaccard) is not much better in terms of throughput than the flat-index baselines seen in Section~\ref{sec:fuzzy-lsh}. Even more surprisingly, this implementation is also poor in terms of recall. The low throughput is caused by Jaccard similarity being significantly more expensive to compute than Hamming distance. While Jaccard similarity requires computations of set intersection and union, Hamming distance can be reduced to bitwise XOR operations.

The explanation for low recall is more subtle. While in the case of Milvus, the low recall was due to truncating the search results, in the FAISS (Jaccard) case the issue stems directly from the similarity metric. At a closer look at edge values inside the graph, we observed that the Jaccard similarity scores can become
tie-heavy. In other words, many vertex pairs receive the same (or nearly the same) similarity score, making the graph harder to navigate reliably.

To further validate this finding, we run a self-search sanity check. After distinct documents from the 100K-document evaluation batch are added to the index, we re-query those same documents. A well-constructed graph should consistently return the query’s own ID near the top of the result list. However, this self-search returns the query's ID for only 4.19\% of recently inserted documents, an unacceptably poor rate for exact matches. 
Together, these results indicate that the Jaccard similarity in its naive implementation is poorly calibrated for graph search, as it tends to treat candidates as either very similar or very dissimilar, with little meaningful structure in between. This approach works when all the signatures in a bucket are compared with each other. However, this technique is too weak for graph exploration. 

\begin{takeawaybox}
\textbf{Takeaway 2.} Graph-index behavior is dominated by the distance metric. The popular similarity metrics for both HNSW and fuzzy deduplication cannot be used out-of-the-box. Hamming distance is fast but misaligned with MinHash/Jaccard deduplication, while naive Jaccard is aligned with the objective but too expensive and tie-heavy for reliable HNSW traversal. A good distance metric must be cheap enough to compute for graph fast traversal while still preserving the neighborhood structure needed for accurate fuzzy-duplicate retrieval.
\end{takeawaybox}

\begin{figure*}
     \centering
    \includegraphics[width=\linewidth]{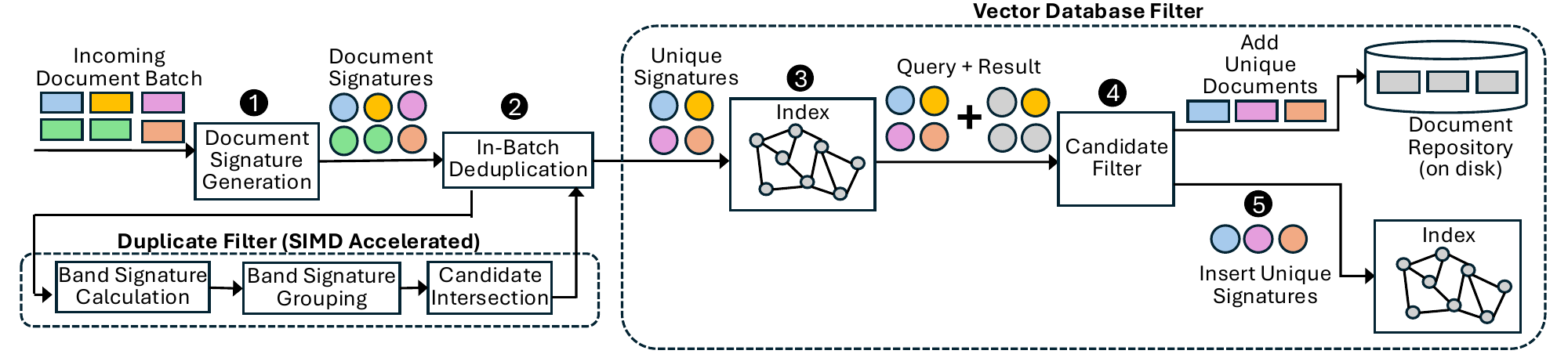}

\caption{\textbf{\papershortname{} workflow.} For each incoming document batch, the documents' bitmap signatures are generated (\ding{202}). Then, \papershortname{} removes near-duplicates inside the batch(\ding{203}). Next, for each input document, the closest neighbors are retrieved from the corpus indexed via an HNSW graph (\ding{204}) and the duplicates are filtered out using a fixed threshold (\ding{205}). Finally, the remaining documents in the input batch are inserted into the corpus and the index, as they are considered unique (\ding{206}).}

     \label{fig:system_workflow}
\end{figure*}

\section{\papershortname{} System Design}
\label{system_design}

 The high-level idea in \( \text{\papershortname}\) is to use an HNSW vector database to quickly detect near-duplicates while supporting updates. To achieve high throughput and recall, we propose an approximation of the Jaccard distance metric using a novel bitmap signature. Together, these optimizations allow for efficient, SIMD-parallelized Jaccard-style distance computation, without sacrificing recall.

\subsection{\papershortname{} Workflow Overview}
\label{sec:rad_flow}

Our first contribution consists of general fuzzy deduplication workflow enabled by using a graph-based vector database. Figure~\ref{fig:system_workflow} shows the end-to-end workflow of \papershortname, consisting of the following high-level steps.

\textbf{(\ding{202}) Document signature generation.} Documents arrive to \papershortname{} in batches.
Each incoming document is shingled and a MinHash signature is generated, then packed into a \papershortname{}-Jaccard bitmap representation (Section~\ref{sec:data_representation}).  From this point onward, \( \text{\papershortname}\)  operates on bitmap signatures.

\textbf{(\ding{203}) Input batch cleanup.}
\papershortname{} removes near-duplicates within each incoming document batch. To speed up deduplication, we apply SIMD acceleration to band processing and candidate intersection computation (Section~\ref{simd_acceleration}). A batch based approach is suitable at this stage, as the incoming batches of documents are assumed to be small relative to the total size of the corpus.

\textbf{(\ding{204}) Index search to retrieve closest neighbors.}
For each document in the clean input batch, \( \text{\papershortname}\)  queries the HNSW index. 
Index Search returns a small set of similar candidates together with their Jaccard-style distances for each input document in the cleaned batch. This step approximates the LSH banding from classic fuzzy deduplication (i.e., by selecting a promising neighborhood of potential near-duplicates), \textit{and} the candidate pair verification (the distance between neighbors is automatically captured by the graph).

 \textbf{(\ding{205}) Input document filtering.} 
 \papershortname{} filters the returned neighbors using the similarity threshold \(\tau\)
(\(\tau \ge 0.7\) in our experiments). For each document in the cleaned input
batch, \papershortname{} scans the returned neighbor list. If any neighbor \(n\) satisfies
\(J_{\mathrm{bm}}(\mathrm{doc}_i,\mathrm{doc}_n)\ge\tau\), equivalently
\(D_{\mathrm{bm}}(\mathrm{doc}_i,\mathrm{doc}_n)\le 1-\tau\), then document
\(i\) is discarded because the corpus already contains a near-duplicate.

\textbf{(\ding{206}) Add unique documents to the corpus.}
Finally, documents that pass the in-batch deduplication and index search filters are considered unique. Their content is written to local \texttt{ext4} files, and their bitmap signatures are inserted into the vector database to support future searches. This update path is central to \papershortname{}: the index is maintained across batches, so future searches operate over the evolving admitted corpus.

\subsection{\papershortname{}-Jaccard: Bitmap-based Signatures}
\label{sec:data_representation}

Our second contribution lies in the data format which enables high throughput and high recall for the workflow presented above. As discussed in Section~\ref{limitations_of_fuzzy_deduplication_frameworks}, an ideal distance function is fast enough to compute for an efficient graph search, and adapted for navigation \textit{in a graph} (i.e., respects the intuition  behind Jaccard similarity, and avoids crowding). We showed that computing the Hamming distance of the MinHash signatures is fast and parallelizable (SIMD- or GPU-friendly), but it is too fine-grained: small, arbitrary changes inside 32-bit signatures can dominate the distance and  distort similarity. In contrast, Jaccard similarity of MinHash signatures matches our deduplication objective, but computing it directly inside a graph index (e.g., FAISS (Jaccard) in Section~\ref{graph_base_comparisson}) is difficult to accelerate with parallelization techniques, and leads to signature crowding. To get the best of both worlds, \papershortname{}-Jaccard derives a bitmap from the original MinHash signatures that enables both SIMD-friendly and accurate search in the HNSW index.

The most challenging hardest part is not only making each comparison fast. The distance metric must also help HNSW build a graph that keeps similar documents close and search that graph accurately under a bounded exploration budget. Intuitively, a document signature and distance metric that achieve high recall need to: 1) preserve the meaning of the distance metric similar to that of
Jaccard similarity to respect the fuzzy deduplication algorithm, and 2) provide enough score separation among near-tied candidates to guide HNSW traversal. This is different from one-shot fuzzy deduplication. There, MinHash--Jaccard is applied to hundreds of thousands or millions of mostly dissimilar documents, so it mainly needs to reject most pairs and flag the few high-overlap candidates. HNSW uses the same score as a navigation signal over a small local neighborhood, with at most \(M=128\) neighbors per node in our evaluation. Since \(\textit{efSearch}\) is fixed to bound read latency as
the graph grows, the score must provide direction, not just rejection.

\begin{figure}[t]
    \centering

            \includegraphics[width=\linewidth]{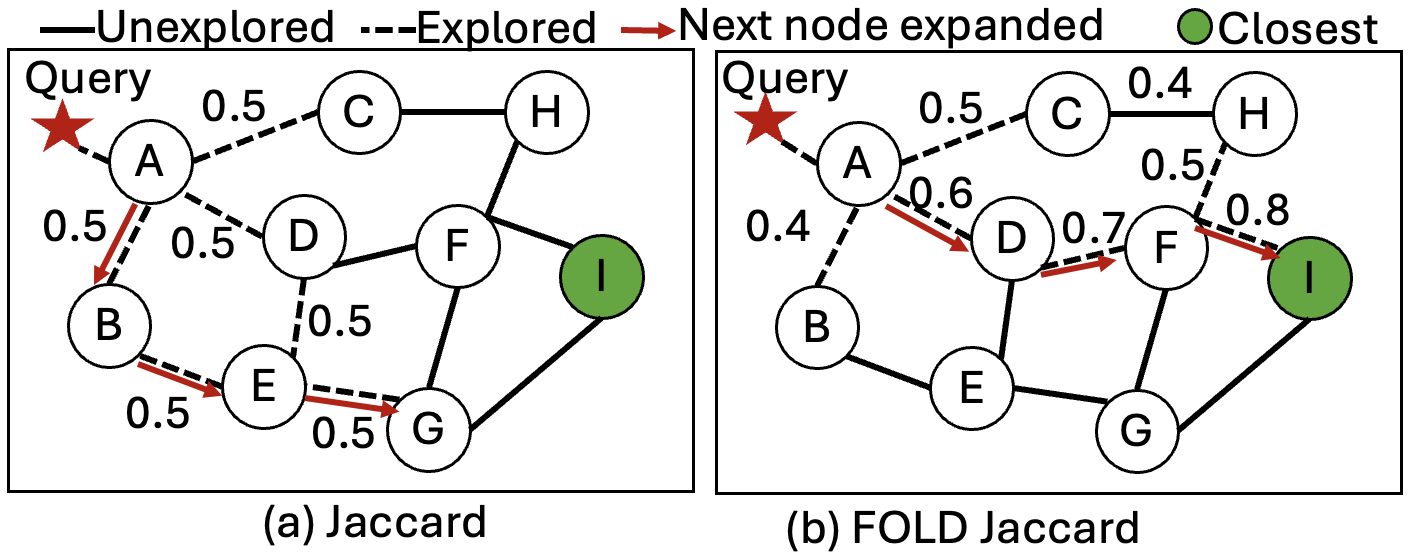}

    \caption{Example HNSW level-0 navigation with $\textit{efSearch}=6$. Search begins at query (star); dashed edges represent neighbor evaluations that consume the $\textit{efSearch}$ budget, and red arrows denote expansion path toward the true closest neighbor $I$ (green). Jaccard (a) cannot break ties, causing breadth-like exploration that fails to reach $I$ within 6-node budget. \papershortname{}-Jaccard (b) separates candidates in bitmap space, giving HNSW clear ordering signal within same budget.}

    \label{fig:hnsw_navigation}
\end{figure}

Breaking ties between similar neighbors is a subtle issue during graph construction. The issue is not that Jaccard is wrong for fuzzy deduplication. It is that raw MinHash--Jaccard is too coarse for graph navigation, because many low-overlap local candidates receive the same low score. We empirically observe that these ties arise in all datasets we analyze when Jaccard similarity is used out-of-the-box during HNSW construction.
Figure~\ref{fig:hnsw_navigation} illustrates this phenomenon. Classic Jaccard scores (Figure~\ref{fig:hnsw_navigation}a) create multiple ties (e.g., $0.5$), forcing exploration to follow breadth-like exploration as it cannot break ties
between neighbors. This would not be an issue if the entire graph was explored. However, in vector databases search is capped to \textit{efSearch} nodes to bound read latency. Hence, search runs out of the exploration budget (using 6 nodes in this example) before reaching the closest neighbor to $Query$, which is $I$. This is especially problematic at the beginning of the search, as the graph walk cannot arrive in the region containing the most similar nodes. As we explain in the rest of this section, \papershortname-Jaccard separates tied candidates, guiding depth-first expansion to reach $I$ more
reliably within the same \textit{efSearch} budget. 

Intuitively, \papershortname{}-Jaccard further groups the search space before computing the similarity, by clustering roughly similar sets of hash functions together. This grouping allows for a more clear direction when exploring the graph, as the input query is guided from cluster to cluster.

\setlength{\abovecaptionskip}{0pt}
\setlength{\belowcaptionskip}{-10pt}
\begin{figure}[t]
    \centering
     
        \includegraphics[width=\columnwidth]{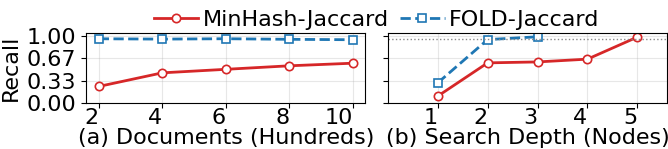}

    \caption{\textbf{Empirical confirmation on a 1,500-document CC-MAIN
    sample with $J\geq 0.7$ ground-truth clusters.}
    (a)~During HNSW construction, \papershortname{}-Jaccard selects \(\approx 95\%\) of
    each node's true top-\(M\) neighbors, compared with \(\approx 51\%\) for
    MinHash--Jaccard.
    (b)~At query time, \papershortname{}-Jaccard reaches recall \(\geq 0.95\) within
    3 opened nodes, while MinHash--Jaccard needs 5 and reaches only
    \(0.70\) by depth 5.}
    \label{fig:bitmap_empirical}
\end{figure}

We validate this effect on real data in Figure~\ref{fig:bitmap_empirical},
using a 1{,}500-document CC-MAIN sample. During construction, we compare the
\(M{=}16\) neighbors selected at each HNSW insertion against the exact
top-\(M\) nearest already-inserted neighbors. \papershortname{}-Jaccard recovers
\(0.95\) of these neighbors, versus \(0.51\) for MinHash--Jaccard. Over 24
held-out queries with \(\textit{efSearch}=K=4\), \papershortname{}-Jaccard
reaches recall \(\geq 0.95\) after 3 opened nodes, while MinHash--Jaccard needs
5; under the depth-5 budget, it reaches only \(0.70\). Thus,
\papershortname{}-Jaccard improves which nodes HNSW explores, not how many.

\begin{figure}
     \centering
         \includegraphics[width=\linewidth]{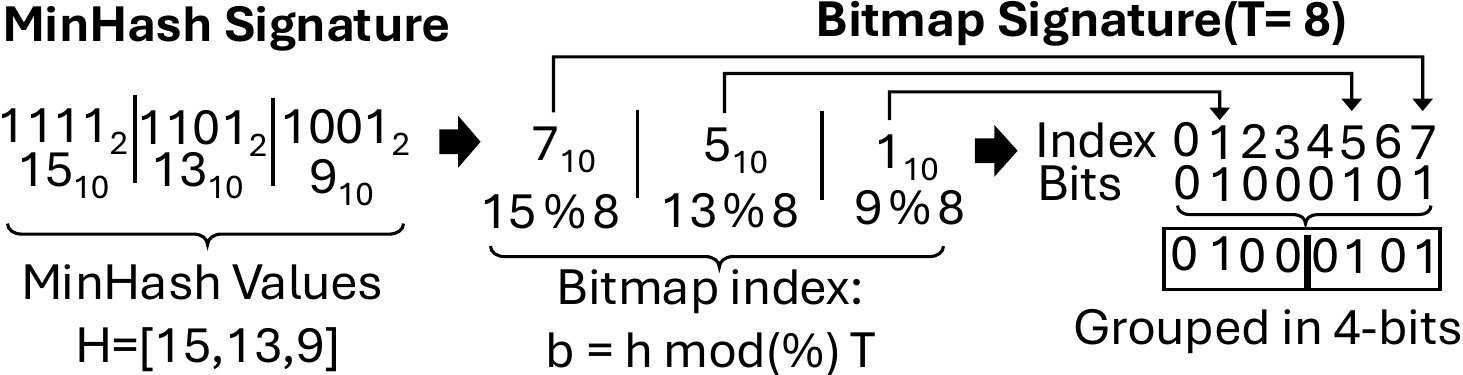}

\caption{\textbf{Example bitmap signature construction from a MinHash signature, with $H{=}3$, $T{=}8$, and 4-bit words.} Each hash value $h$ maps to an index $b = h \bmod T$, and the bitmap sets $x[b] \leftarrow 1$. \papershortname\ uses $T{=}4096$ and $H{=}112$ MinHash values, packed into 32- or 64-bit words.}

     \label{fig:two_level_sketch_underbrace_compact}
\end{figure}

Figure~\ref{fig:two_level_sketch_underbrace_compact} shows a small example of how bitmap signatures are created in \papershortname{}. First, shingling and MinHashing  produce  a fixed-length vector of \(H\) MinHash values. We illustrate $H=3$ MinHashes: $15, 13,$ and $9$ (in practice, $H=112$). \papershortname's bitmap signature has size \(T\) bits. $T = 8$ in our example (in practice, $T = 4096$), where each bit corresponds to a position that will be "turned on" by each of the MinHashes. The bitmap is initialized to all zeros. Then, each MinHash value \(h\) is mapped to a bitmap position \(h \bmod T\). In the example, positions 1, 5, and 7 are turned on, to create the final bitmap signature $[0100\ 0101]$ (note that collisions are possible).

\setlength{\abovecaptionskip}{0pt}   
\setlength{\belowcaptionskip}{-10pt} 

This representation allows for the computation of the intersection and union of two MinHash signature sets using bitwise operations. Given two signatures $A$ and $B$, the intersection is represented by the number of common positions that are set to 1 in both $A$ and $B$. The union is represented by the total number of 1 bits in $A$ and $B$. Note the Jaccard similarity approximation on the bitmaps stays close the meaning of the Jaccard similarity: if $A$ and $B$ have the same MinHash in their initial signature set, then they will both "turn on" the same bit in the bitmap signatures. The total number of turned on bits across both bitmaps approximate the total number of distinct MinHash signatures for $A$ and $B$.

Though this approximation of the Jaccard similarity is amenable to parallelization, it also introduces the risk of collisions. This can arise from two sources: (1) two MinHashes in separate signatures accidentally turn on the same bit in the bitmap, and (2) two MinHashes in the same signature accidentally turn on the same bit in the bitmap.
Maybe counter-intuitively, these collisions can help the search advance faster in the beginning of the exploration by reducing score crowding. Consider the following example, consisting of an incoming query $Q$, with its MinHash signature $[9,13,15,18,22,27]$, and two neighbor documents $A$ and $B$, with MinHash signatures $[9,13,15,18,14,28]$ and $[9,13,15,18,16,28]$, respectively. We consider $T=8$, as above. The MinHash-Jaccard similarity gives $J(Q,A)=J(Q,B)=0.5$---a tie. However, after folding the signatures into bitmaps, the bitmap-level score separates the pair: $J_{\text{bitmap}}(Q,A)=0.71$ whereas $J_{\text{bitmap}}(Q,B)=0.5$.

In the above example, $A$ is not necessarily a closer  neighbor than $B$. Since the example is small, the variation in the bitmap Jaccard similarity scores is significant. However, if $T$ is large enough to stay close enough to the Jaccard similarity value despite a few collisions (as we show below), then the collisions are useful to effectively break ties in the case of  neighbors that are very similar and that would obtain the same Jaccard similarity value. Effectively, these collisions give \papershortname{} a stronger ordering signal during bounded HNSW traversal, making the search less  breadth-first exploration  and helping it reach more promising neighbors earlier. This is important, especially at the beginning of the search, as HNSW graphs have an exploration budget (determined by the \textit{efSearch} parameter in the FAISS implementation).

\noindent\textbf{Collision analysis for bitmap signatures.}
We provide a high-level analysis of why bitmap collisions do not significantly distort Jaccard-aligned scores at our operating scale, while still reducing score crowding during HNSW traversal. Appendix~\ref{appendix:bitmap_collision_analysis} provides the full proof.

\papershortname{} works with 4096-bit bitmaps (\(T=4096\)) and 112 MinHash signatures. The analysis proceeds in two steps: within one document, we estimate how many distinct bits are on and how many hashes collide; across unrelated documents, we estimate how much overlap occurs by chance and whether it can exceed our deduplication threshold. For \(T=4096\), a document has \(s\approx110.50\) active bits on average, so only \(\approx1.50\) of the 112 hashes collide within a document. Two unrelated documents share only about three active bits in expectation, giving a typical \papershortname{}-Jaccard score around \(0.014\). In contrast, satisfying \(J_{\text{bitmap}}\ge0.7\) requires roughly \(91\) shared bits; under the corresponding hypergeometric model, the probability of such accidental overlap is \(\approx5.95\times10^{-147}\). Thus, random bitmap collisions are extremely unlikely to create false positives at our operating threshold.

\section{Implementation and Optimizations}
\label{implementation}

\papershortname{} is a multi-threaded C++/Python system built on FAISS HNSW. Our implementation runs on CPUs, but the design could be incorporated into a GPU implementation. C++ implements the hot-path distance kernels and bitmap primitives, while Python orchestrates ingestion and batch deduplication.
We describe the implementation choices that sustain high-throughput streaming deduplication in \papershortname{}: SIMD acceleration for Jaccard distance computation (Section~\ref{simd_acceleration}), and caching to reduce repeated work in HNSW search/construction hot loops (Section~\ref{sec:vector_database_filter}).

\subsection{SIMD Acceleration of Jaccard Similarity}
\label{simd_acceleration}

Modern processors provide SIMD extensions that apply one operation to multiple data elements in parallel, accelerating database and vector-search workloads~\cite{DBLP:journals/pvldb/LarsonBHHNP15,10.1145/2723372.2747645,faiss_git_docs,DBLP:series/synthesis/2015Hughes}. \papershortname{} applies SIMD optimizations in two parts of the overall workflow. First, as shown in Figure~\ref{fig:system_workflow}, \papershortname{} accelerates the input-batch deduplication, which follows the classic fuzzy deduplication flow (described in Section~\ref{fuzzy_deduplication_background}). The band calculations as well  as the candidate intersection are accelerated as follows. Recall that a MinHash signature is an array of size \(H\), consisting of 32-bit values. Given two signatures, the MinHash-Jaccard estimate is the fraction of positions (lanes) where the two 32-bit values are identical. SIMD accelerates this by comparing multiple 32-bit lanes at once and using a bitmask and \texttt{popcount} to count matches.

Second, \papershortname{} applies SIMD optimizations to efficiently compute the \papershortname{}-Jaccard similarity when querying the graph index. As described in Section~\ref{sec:data_representation}, each MinHash signature is mapped to a sparse bitmap of length \(T=4096\). For two bitmaps \(A,B\), the \papershortname{}-Jaccard similarity depends only on three \texttt{popcounts} (i.e., counting the number of 1 bits): $p_a = \mathrm{popcount}(A)$,
$p_b = \mathrm{popcount}(B)$,
and $p_x = \mathrm{popcount}(A \oplus B)$. Since \(p_a+p_b = 2|A\cap B| + p_x\), it follows that the intersection $I = (p_a + p_b - p_x)/2$, the union $U = (p_a + p_b + p_x)/2$, and the Jaccard similarity $J=I/U$. The corresponding distance used by HNSW is therefore \(D = 1-J = 2p_x/(p_a+p_b+p_x)\), matching Algorithm~1.

Bitmaps are stored as \(W=T/64\) 64-bit machine words (for \(T=4096\),
\(W=64\)). SIMD accelerates the three popcounts by processing multiple
words per iteration: we load a block of words from \(A\) and \(B\), compute
word-wise XOR for \(A\oplus B\), apply vector popcount to each stream, and
accumulate the partial sums. After scanning all \(W\) words we obtain
\(p_a=\mathrm{popcount}(A)\), \(p_b=\mathrm{popcount}(B)\), and
\(p_x=\mathrm{popcount}(A\oplus B)\), and compute \(I,U,J\) with the scalar formulas above. Thus per-pair scoring reduces to a loop of vector loads, XOR, vector-popcount, and a few scalar additions.

\subsection{Caching Optimizations for the Graph Index}
\label{sec:vector_database_filter}

To accelerate index search, \papershortname{} caches the \texttt{popcount} values of stored bitmap signatures.
At query time, \papershortname{}'s search routine uses the bitmap signature of the current query \(A\) together with cached \texttt{popcounts} for each traversed node: the query \texttt{popcount} \(p_a=\mathrm{popcount}(A)\), computed once per query, and a per-vector array \(\{p_b[i]\}\), where \(p_b[i]=\mathrm{popcount}(B_i)\).

\begin{algorithm}[hb]
  \caption{Jaccard distance between $A$ and neighbor $B_i$}
  \label{alg:\papershortname{}-Jaccard}
  \begin{algorithmic}[1]
\Require query bitmap $A$, cached \(p_a=\mathrm{popcount}(A)\), cached \(p_b[i]=\mathrm{popcount}(B_i)\).
    \State $p_x \gets \mathrm{popcnt}(A \oplus B_i)$
      \Comment{SIMD XOR + popcount}
      \label{alg:line-px}
    \State \Return $D(A,B_i)=2p_x/(p_a+p_b[i]+p_x)$
      \label{alg:line-ret}
  \end{algorithmic}
\end{algorithm}

\papershortname{} maintains a per-vector array \(\{p_b[i]\}\) with one entry per bitmap \(B_i\). When new nodes are added, \papershortname{} computes $p_b[i] = $ \texttt{popcount}$(B_i)$ once and stores it in a 16-bit slot (\texttt{uint16}), adding 2~bytes of metadata per stored vector.
During graph search, each visited neighbor \(B_i\) is scored using Algorithm~\ref{alg:\papershortname{}-Jaccard}. This routine is invoked for every visited neighbor, and, due to caching, requires a single 4096  bit SIMD \texttt{XOR+popcount} operation: Line~\ref{alg:line-px} computes \(p_x = \mathrm{popcount}(A \oplus B_i)\). The remaining work uses cached data.

During graph construction, \papershortname\ must also measure distances between two nodes \(B_i\) and \(B_j\) corresponding to documents \(i\) and \(j\). \papershortname{} computes \(p_x = \mathrm{popcnt}(B_i \oplus B_j)\) with the SIMD kernel, reads \(p_b[i]\) and \(p_b[j]\) from the precomputed array, and returns \(D(i,j) = D(p_b[i], p_b[j], p_x) =\)
\(2p_x/(p_b[i] + p_b[j] + p_x)\). Thus, on both the query and construction paths, the dominant cost per comparison is reduced to a single 4096-bit \texttt{XOR+popcount}. Caching the query popcount once per query and precomputing \(p_b\) once per document eliminates redundant work in the hottest loops, while adding only 2~bytes of metadata per stored vector.

\subsection{Parameter Tuning and Scalability}
\label{sec:rad_scalability_discussion}

\papershortname{} uses one fixed HNSW configuration in the evaluation. This makes the early-scale points conservative: the same \(M\) and \(\texttt{efSearch}\) budget chosen for the largest and noisiest workloads is also used when the index is small. This matters for streaming deduplication because missed neighbor admit duplicates into the growing corpus. A natural next step is an adaptive policy that monitors retrieval quality,
adjusts \(\texttt{efSearch}\) during search and \(\texttt{efConstruction}\) for new insertions, and refreshes or rebuilds the index with a larger \(M\) when recall degrades. Longer-running streams may also need disk-backed ANN storage. Parallelization is complementary: all systems can benefit from sharding or hardware acceleration, and \papershortname{} can shard HNSW indexes, merge per-shard candidates, and parallelize \papershortname{}-Jaccard scoring across GPUs~\cite{harmony,gottesburen2024unleashing}. These
optimizations improve absolute throughput, but do not change core comparison evaluated next: \papershortname{} bounds per-document search while preserving Jaccard-aligned retrieval signal under continuous growth.

In our implementation, larger values of $M$,$\texttt{efSearch}$ and $\texttt{efConstruction}$ consistently yielded empirical recall $=1.0$ for \papershortname{} on our datasets, but not for the FAISS (Jaccard) baseline. Because these larger settings also increased index construction and query cost, we use $M = 128$, $\texttt{efConstruction} = 512$, $\texttt{efSearch} = 400$, and $k=4$ as a balanced operating point. Smaller values were sufficient for cleaner, partially deduplicated corpora such as C4, but degraded recall on noisier, highly duplicated datasets such as Common Crawl.
\section{Experimental Evaluation}
\label{experimental_evaluation}

We evaluate \papershortname{} in a setting where documents arrive continuously and need to be deduplicated against the existing corpus, which grows over time. We answer the following:

\begin{enumerate}[leftmargin=*, nosep, noitemsep]
    \item \textbf{End-to-end throughput and recall:} How does
    \papershortname{} compare with DPK~\cite{ibmdpk},
    Milvus~\cite{milvus}, and FAISS (Jaccard) across diverse
    real-world datasets? (Section~\ref{end_to_end_throughput_latency_correctness})

    \item \textbf{Performance breakdown:} Which internal components
    dominate runtime in \papershortname{}?
    (Section~\ref{performance_breakdown})

    \item \textbf{Scalability with dataset size:} Can
    \papershortname{} maintain stable throughput as the corpus grows?
    (Section~\ref{scalability})
\end{enumerate}

\subsection{Experimental Setup}
\label{experiment_setup}

\noindent\textbf{Datasets.}
We evaluate four corpora commonly used for LLM training and prior deduplication studies~\cite{guo2020wiki,chelba2013one, 10.5555/3455716.3455856,DBLP:journals/corr/abs-2104-08758,
10.5555/3454287.3455099,lee-etal-2022-deduplicating,Zhang2023RETSimRA}: LM1B (30.3M documents), RealNews (32.8M documents), and 30M-document samples from C4 and a recent
Common Crawl snapshot. Table~\ref{tab:dataset_text_stats} summarizes their redundancy, document-length distribution, and shingle volume. Importantly, Table~\ref{tab:dataset_text_stats} shows that the datasets capture different duplicate proportions. Duplicate counts are obtained using the DPK fuzzy-deduplication pipeline, at the standard fuzzy-deduplication threshold $J \ge 0.7$. As we explain in Section~\ref{limitations_of_fuzzy_deduplication_frameworks}, computing the ground truth using a brute-force method is prohibitively expensive in terms of time. For a 3M document dataset, brute-force requires several days with our available hardware. Given that our evaluation focuses on scalability (with 10x larger datasets), and that the cost to brute-force is quadratic in the number of documents, we compare the recall against DPK, which is an industry-standard representative of the LSH deduplication pipeline. We validate DPK against brute force on 3M-document subsets and find that DPK reaches 0.92 recall against brute-force (Section~\ref{limitations_of_fuzzy_deduplication_frameworks}).

\begin{table}[h]
  \scriptsize
  \setlength{\tabcolsep}{3pt}
  \renewcommand{\arraystretch}{0.95}
  \resizebox{\columnwidth}{!}{%
    \begin{tabular}{@{}lrrrr@{}}
      \toprule
      Dataset & Documents(M) & Duplicates & p99w & shingle5(B) \\
      \midrule
      LM1B     & 30.30 & 601{,}554 (1.98\%)        & 64    & 0.65 \\
      RealNews & 32.80 & 2{,}364{,}644 (7.20\%)    & 2{,}505 & 18.78 \\
      C4       & 30.00 & 608{,}791 (2.02\%)        & 2{,}675 & 10.74 \\
      Common Crawl & 30.00 & 12{,}199{,}957 (40.66\%) & 6{,}683 & 28.66 \\
      \bottomrule
    \end{tabular}%
  }

\caption{\textbf{Workload diversity across corpora.} Datasets range from short, low-redundancy text to long, highly redundant raw-web documents. Duplicates are detected by DPK at $J \ge 0.7$; p99w is the 99th-percentile length in words, and shingle5 is the number of 5-word shingles in billions.}

  \label{tab:dataset_text_stats}
\end{table}

\noindent\textbf{Baselines.}
Section~\ref{limitations_of_fuzzy_deduplication_frameworks} identifies the main baseline design points and their limitations under continuous ingestion. For the full-scale evaluation, we carry forward the three most relevant systems: DPK~\cite{ibmdpk}, Milvus~\cite{milvus}, and FAISS (Jaccard). DPK provides the recall reference and industry-standard for one-shot deduplication; Milvus represents the production vector-database baseline; and FAISS (Jaccard) is a baseline we implement to isolate the effect of using HNSW with the out-of-the-box Jaccard similarity score.

\noindent\textbf{System configuration.}
Unless otherwise specified, \papershortname{} and FAISS (Jaccard) use the same HNSW configuration: $M=128$, $\texttt{efConstruction}=512$, and $\texttt{efSearch}=400$. This gives both graph-based methods the same search and construction budget, so recall and throughput differences reflect the representation and distance-computation path rather than parameter choices. Both methods return $k=4$ candidates per query. All methods use the same near-duplicate threshold, $J \ge 0.7$, following common practice in production fuzzy deduplication frameworks~\cite{datajuicer,redpajama,ibmdpk}.

\noindent\textbf{Hardware.}
The configuration is the same as in  Section~\ref{limitations_of_fuzzy_deduplication_frameworks}.

\noindent\textbf{Methodology.}
We evaluate continuous ingestion with repeated 1M-document growth cycles until each workload reaches 30M documents. Each cycle uses the same measurement structure as Section~\ref{limitations_of_fuzzy_deduplication_frameworks}, but at a larger scale: 900K documents are used to grow the corpus and index, and the  following 100K documents form the evaluation slice. We report recall and throughput on these 100K-document slices.

\subsection{End-to-end Throughput and Recall}
\label{end_to_end_throughput_latency_correctness}

\begin{figure*}
     \centering
        \includegraphics[width=\linewidth]{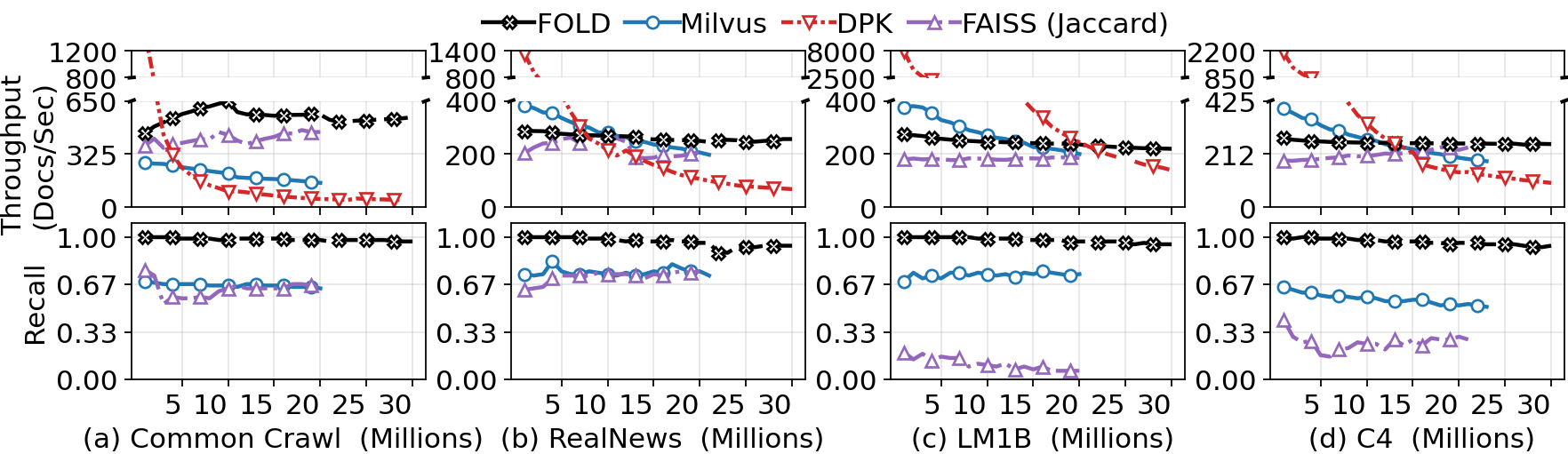}

\caption{\textbf{\papershortname{} preserves the high-throughput, high-recall operating point as the corpus grows.} 
At 30M documents, \papershortname{} maintains 0.94--0.97 recall and 220--551 docs/sec across the four workloads. DPK and Milvus lose throughput as the corpus grows, while FAISS (Jaccard) keeps bounded graph-search throughput but its recall remains lower and dataset-dependent. }

     \label{fig:throughput_recall_main_figure}
\end{figure*}

Figure~\ref{fig:throughput_recall_main_figure} reports end-to-end throughput and recall as each workload grows from 10M to 30M documents. The key observation is the scaling trajectory: an online deduplication system must keep processing new documents quickly after the index has grown, while still finding near-duplicates accurately. A method that is fast only early in the run, or fast because it misses duplicates, does not meet fuzzy deduplication scalability requirements.

\papershortname{} is the only system that maintains both high throughput and high accuracy across all datasets. At the largest corpus size (30M), its recall remains high: 0.94 on C4, 0.97 on Common Crawl, 0.95 on LM1B, and 0.94 on RealNews. Common Crawl is the most difficult case (with the most duplicates, ~40\%), and also the case where RAD shines. The throughput increases from 454 to 551 docs/sec while recall remains high, at 0.97. As the corpus grows, \papershortname{} filters more incoming documents before they are inserted. Those dropped documents never enter the HNSW index, so later cycles pay less insertion and index-maintenance cost. 
In the lower-redundancy workloads, \papershortname{} keeps throughput nearly flat; in the high-redundancy workload, accurate filtering creates an additional benefit by slowing index growth. On C4, throughput moves from 277 to 253 docs/sec. RealNews and LM1B follow the same pattern of relatively flat throughput, as the working dataset size increases. 

The other baselines cannot maintain \textit{both} high throughput and high recall as the corpus size grows. DPK shows the expected scaling bottleneck of batch-oriented fuzzy deduplication. Its throughput falls by up to 50x (from 7{,}823 to 139 docs/sec on LM1B) with the throughput decrease in the other datasets being 19x, on average. The issue is that each incoming document batch is processed against an ever-larger accumulated corpus, so candidate generation and verification cost eventually dominate the insertion path.

Milvus does not preserve either throughput or recall at  30M documents. Its recall remains well below \papershortname{}, and not good enough to maintain a high-quality dataset as the corpus grows. On C4, recall reaches 0.47 (similar to a random decision). Even in the best-performing case (RealNews), Milvus' recall stays low, at 0.75. Its throughput also falls by up to 4x,  as the datasets scale, staying, on average 2.8x  lower than \papershortname.  This is the same throughput-recall tension observed earlier: a small candidate budget keeps retrieval cheaper but misses duplicates, while a larger candidate set would increase Jaccard similarity computation work.
Finally, FAISS (Jaccard) keeps its throughput steady as the corpus grows. However, this speed does not come together with reliable deduplication. Recall falls and is dataset dependent:  C4 recall drops to 0.27, Common Crawl to 0.66,  LM1B to just  0.10, and RealNews to 0.82. Thus, graph search alone solves only the throughput side of the problem. Without a stronger Jaccard-aligned retrieval signal, search can stay fast while still missing too many duplicate neighborhoods.

\subsection{Performance breakdown}
\label{performance_breakdown}

\begin{figure*}[h]
     \centering
    \includegraphics[width=\linewidth]{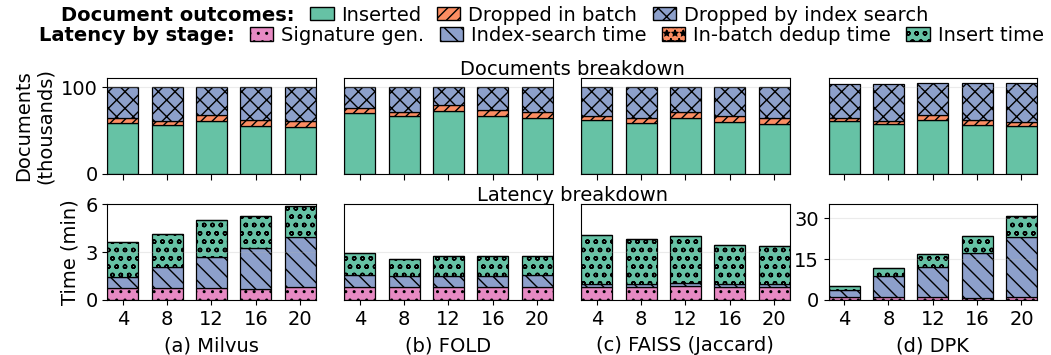}

\caption{\textbf{Common Crawl breakdown.}
Top: document outcomes per 100K-document streaming slice. Bottom: latency by stage. \papershortname{} keeps search time stable and turns duplicate drops into lower insertion time; Milvus and DPK lose those savings to growing search cost, while FAISS (Jaccard) stays low-latency but low-recall. All methods are shown to 4M-20M documents.}

     \label{fig:latency_breakdown_cc_main}
\end{figure*}

Figure~\ref{fig:latency_breakdown_cc_main} provides a breakdown of the processed document outcomes (top) and latency (bottom). We focus on Common Crawl because it is the most difficult workload, which stresses both duplicate filtering and index growth. The breakdown is shown up to 20M documents for readability. For each 100K-document streaming slice, the top row shows how many documents are inserted or dropped, and the bottom row shows where time is spent. In-batch deduplication is visually negligible at this scale, ranging from 0.056--0.158\,s across methods. The important costs are therefore index search and insertion.

For \papershortname{}, the breakdown shows the desired behavior for a streaming system. Across the plotted 4M--20M range, documents dropped by index search rise by 18\% (24{,}465$\rightarrow$28{,}829), so insertions fall by 9\% (70{,}443$\rightarrow$64{,}140). Signature generation and index search remain nearly flat, moving from 0.81$\rightarrow$0.80\,min and 0.72$\rightarrow$0.73\,min, respectively. The main change is insertion time: as fewer documents enter the index, insert time falls by about 13\%, from 1.38$\rightarrow$1.20\,min. This lowers total latency by 6\%, from 2.91$\rightarrow$2.74\,min. Since Common Crawl has high duplicate pressure, accurate duplicate filtering helps twice: it removes redundant documents and
reduces future index-maintenance work.

Milvus also inserts fewer documents as redundancy increases, but it does not
keep search cost under control. Across the plotted 4M-20M range, documents
dropped during index search rise by about 9\% (35{,}849$\rightarrow$39{,}124), while insertions fall by about9\% (59{,}059$\rightarrow$53{,}845). This reduces insert time by about 11\%, from 2.19$\rightarrow$1.96\,min. However, candidate-retrieval time grows by 4.5$\times$, from 0.69$\rightarrow$3.14\,min, as the indexed corpus grows. As a result, total latency rises by about  64\%, from 3.60$\rightarrow$5.91\,min. Thus, Milvus gets some insertion savings, but the growing search cost more than erases them.

DPK shows the scaling bottleneck most sharply. Across the plotted 4M-20M range, documents dropped during search rise by about 15\% (39{,}403$\rightarrow$45{,}244), while inserted documents fall by about 10\% (60{,}597$\rightarrow$54{,}756). Search time grows by 8.5$\times$ (2.62$\rightarrow$22.24\,min), and insert/update time grows by 4.8$\times$ (1.58$\rightarrow$7.52\,min). As a result, total latency rises by
6.1$\times$, from 5.01$\rightarrow$30.57\,min. This matches the earlier throughput trend: as the corpus grows, later
streaming slices become costlier.

FAISS (Jaccard) has a different profile. Across the plotted 4M--20M range, index search time remains small, increasing slightly from 0.17$\rightarrow$0.21\,min. It also drops more documents at larger scales: documents dropped during index search   rise by about 8\%
(33{,}132$\rightarrow$35{,}701), while insertions fall by  about 7\% (61{,}776$\rightarrow$57{,}268). This reduces insert time by about 23\%, from 3.08$\rightarrow$2.36\,min, and lowers total latency by about 17\%,  from 4.06$\rightarrow$3.37\,min. However, this low-latency profile comes with the low recall seen in Section~\ref{end_to_end_throughput_latency_correctness}.
Graph search  keeps the work bounded, but raw MinHash--Jaccard scoring does
not reliably guide HNSW to the right duplicate neighborhoods. We confirm this retrieval-quality gap with a self-search diagnostic, similar to Section~\ref{limitations_of_fuzzy_deduplication_frameworks}. After inserting a 100K-document batch, we query those same documents against the index. \papershortname{} returns the query's own ID in the top-$k$ list for 98.7\% of the documents, while FAISS (Jaccard) does so for only 16.8\%. Thus, FAISS's low latency is not an accuracy advantage: its graph search is cheap, but its local neighborhoods are often wrong.

\begin{figure}[t]
    \centering

        \includegraphics[width=\columnwidth]{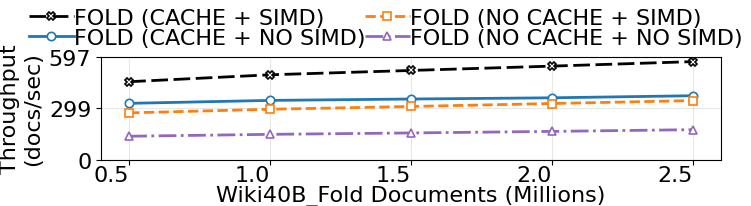}
   
\caption{\textbf{Performance breakdown of \papershortname{} optimizations.} All \papershortname{} variants use the HNSW index and bitmap signatures; only popcount caching and SIMD are toggled.}
    \label{fig:\papershortname_ablation_cc_main}
\end{figure}

Figure~\ref{fig:\papershortname_ablation_cc_main} isolates the distance computation used inside \papershortname{}'s HNSW traversal. The four curves differ only in the   distance implementation: whether popcount values are cached  (CACHE + NO SIMD), whether SIMD is used (NO CACHE + SIMD), or both (CACHE + SIMD). The index, bitmap signatures, and retrieval settings are unchanged. 

Figure~\ref{fig:\papershortname_ablation_cc_main} reports throughput only, as recall remains 1.00 for all variants. The  throughput differences come from differences in the speed of distance computation rather than  changes in retrieval quality. At the largest plotted scale, the scalar baseline without caching or SIMD reaches 176 docs/sec. SIMD alone increases throughput to 344 docs/sec, while caching alone reaches 372 docs/sec by avoiding repeated popcount work. Combining both optimizations gives the default \papershortname{} configuration, which reaches 569 docs/sec, a 3.3$\times$ improvement over scalar on-the-fly computation at the same recall.

\subsection{Scalability with Dataset Size}
\label{scalability}

We finally test whether the throughput trend from Figure~\ref{fig:throughput_recall_main_figure} holds beyond the 30M-document evaluation by extending Common Crawl to 50M documents. We use the same 1M-document ingestion cycle from Section~\ref{experiment_setup} and measure end-to-end throughput on each 100K-document streaming slice as the indexed corpus grows.

\begin{figure}[t]
    \centering
        \includegraphics[width=\columnwidth]{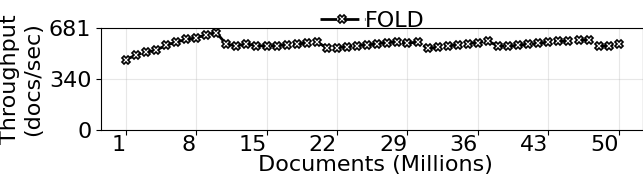}

    \caption{\textbf{\papershortname{} sustains throughput through 50M Common Crawl documents.} The throughput remains steady, between 544--599 docs/sec as the dataset scales, with no late-scale throughput collapse under continuous insertion.}
    
    \label{fig:probe_throughput_export_throughput}
\end{figure}

Figure~\ref{fig:probe_throughput_export_throughput} shows that \papershortname{} remains stable across the full 1M--50M range. In this extended run, throughput starts at 467 docs/sec at 1M, peaks at 648 docs/sec at 10M, and then stays in a narrow 544--599 docs/sec band from 11M to 50M, ending at 574 docs/sec. The key result is the steady state: \papershortname{} does not show a late-scale throughput collapse under continuous insertion. Its candidate retrieval, bitmap--Jaccard scoring, and index updates remain efficient even when the HNSW index reaches tens of millions of documents.

\section{Related Work}
\label{rw}

Data deduplication is important in LLM pre-training because redundant data can
increase training time, reduce generalization, and amplify
memorization~\cite{hernandez2022scaling,DBLP:conf/iclr/CarliniIJLTZ23}.
More broadly, deduplication removes repeated content at different levels.
Storage deduplication removes repeated byte chunks or segments to reduce
capacity, backup/restore cost, and indexing overhead~\cite{paulo2014survey,
zhu2008avoiding,lillibridge2009sparse,srinivasan2012idedup,zou2022mega,
pan2025gogetafs}. These exact fingerprint based techniques are complementary
to text deduplication: documents can remain near-duplicates under
shingle-based  Jaccard similarity even when edits, formatting changes,
boilerplate, or reordering alter their byte-level chunks.

Text deduplication can be exact or fuzzy. Exact text deduplication removes
identical documents or repeated substrings, using methods such as document
hashing or suffix-array-based approaches~\cite{doi:10.1137/0222058}. Such
methods are useful for exact copies, but they miss many near-duplicates that
differ lexically while still containing substantially overlapping content.
Fuzzy text-deduplication approaches target these cases and commonly rely on
sketches such as MinHash~\cite{666900} and
SimHash~\cite{10.1145/509907.509965}, LSH-style candidate generation,
prefix-filter set-similarity joins, and related set-search
systems~\cite{xiao2008efficient,xiao2009topk,xiao2011efficient,
vernica2010efficient,zhu2016lshensemble,zhu2019josie,
fernandez2019lazo,esmailoghli2022mate}. These ideas appear in LLM
data-curation systems such as IBM DPK~\cite{ibmdpk},
Data Juicer~\cite{datajuicer}, DataTrove~\cite{datatrove},
RedPajama~\cite{redpajama}, NeMo-Curator~\cite{nvidiadatacuration},
and Milvus~\cite{milvus}. \papershortname{} targets the same fuzzy lexical
deduplication objective, but changes the retrieval path: instead of relying
on growing LSH buckets, prefix-filter candidate sets, or flat candidate
retrieval, it maintains an incrementally updated HNSW index over admitted
documents and retrieves a bounded candidate neighborhood for each incoming
document.

Graph-based ANN indexes such as HNSW are widely used for vector
search~\cite{hnsw_faiss}. \papershortname{} does not propose a new ANN index;
rather, it shows how to make graph search work for online fuzzy
deduplication. Off-the-shelf HNSW distance signals are not sufficient for
MinHash/Jaccard deduplication: Hamming distance is fast but misaligned, while
raw MinHash--Jaccard is aligned but expensive and tie-heavy inside graph
traversal. \papershortname{} addresses this gap with bitmap--Jaccard scoring,
which preserves the lexical Jaccard signal while making bounded graph search
discriminative and cheap enough for continuous ingestion.

Semantic deduplication uses pre-trained embeddings and vector databases to
identify documents with similar meaning~\cite{abbas2023semdedup}. This
optimizes embedding-space similarity, whereas \papershortname{} targets
syntactic near-duplicates under shingle-based Jaccard similarity, the
objective used by MinHash/LSH fuzzy-deduplication pipelines. Thus,
\papershortname{} complements semantic deduplication rather than replacing
it. Overall, \papershortname{} addresses the scalability gap in fuzzy text
deduplication: it preserves the lexical Jaccard objective while
making candidate retrieval fast enough for continuously evolving corpora.

\section{Conclusion}
\label{conclusion}

We introduced \papershortname{}, an online fuzzy deduplication system for continuously growing LLM corpora. \papershortname{} replaces repeated global bucket construction and growing candidate scans with bounded HNSW search over admitted documents, and uses bitmap-Jaccard signatures to make that search both Jaccard-aligned and cheap to compute. Across LM1B, C4, RealNews, and Common Crawl, \papershortname{} preserves  high recall with stable end-to-end throughput as the corpus grows. At 30M documents, it maintains 0.94--0.97 recall and is faster than DPK and Milvus on every workload. These results show that fuzzy deduplication can remain both accurate and scalable under continuous data ingestion.

\clearpage

\enlargethispage{-3\baselineskip}

\bibliographystyle{plainnat}
\bibliography{references}
\appendix

\section{Appendix: Bitmap Collision Analysis}
\label{appendix:bitmap_collision_analysis}

\noindent\textbf{Overview.}
We provide the full derivation for the collision analysis summarized in the main paper. The goal is to show why bitmap collisions do not significantly affect the Jaccard similarity scores between nodes, while reducing score crowding during HNSW traversal. \papershortname{} works with 4096-bit bitmaps (\(T=4096\)) and 112 MinHash signatures.  We proceed in two steps: (1) within one document, how many distinct bits do we expect to be on and how many hashes collide; and (2) across two unrelated documents, how much overlap should we expect purely by chance, and whether that overlap could realistically exceed our deduplication threshold.

\paragraph{Step 1: Expected distinct 1-bits and within-document collisions.}
We model folding \(H\) MinHash values into a \(T\)-bit bitmap as a balls-into-bins process: each of the \(H\) values maps independently and uniformly to one of the \(T\) bit positions.
For any fixed bit position \(t\), the probability that none of the \(H\) values land on \(t\) is \((1-\tfrac{1}{T})^{H}\), so
\[
P(\text{bit OFF}) = \left(1 - \frac{1}{T}\right)^{H},\qquad
P(\text{bit ON}) = 1 - \left(1 - \frac{1}{T}\right)^{H}.
\]
By linearity of expectation, the expected number of distinct 1-bits per document is
\[
s \;=\; \mathbb{E}[\#\text{ones}]
\;=\; T\,P(\text{bit ON})
\;=\; T\Bigl(1 - \bigl(1-\tfrac{1}{T}\bigr)^{H}\Bigr),
\]
and the expected number of within-document collisions is
\[
\mathbb{E}[\text{collisions}] = H - s.
\]

Table~\ref{tab:bitmap_collisions_appendix} reports these values for \(H=112\) and several bitmap sizes \(T\).
For \(T=4096\), we get \(s \approx 110.50\), i.e., only \(\approx 1.50\) of the 112 hashes collide on average. This keeps the bitmap footprint close to the original Jaccard similarity. Larger \(T\) reduces collisions further, but with diminishing returns and higher memory cost.

\begin{table}[h]
  \centering
  \caption{Expected distinct bits and collisions when mapping \(H=112\) MinHash values into a bitmap of size \(T\).}
  \label{tab:bitmap_collisions_appendix}
  \setlength{\tabcolsep}{3pt}
  \begin{tabular}{rccc}
    \toprule
    \(T\) (bits) & \(T/8\) (bytes) & \(\mathbb{E}[\#\text{ones}]=s\) & \(\mathbb{E}[\text{collisions}]=H-s\) \\
    \midrule
    2{,}048 & 256     & 109.02  & 2.98 \\
    4{,}096 & 512     & 110.50  & 1.50 \\
    8{,}192 & 1{,}024 & 111.24  & 0.76 \\
    \bottomrule
  \end{tabular}
\end{table}
\paragraph{Step 2: Accidental overlap between unrelated documents.}
After folding, each document becomes a set of active bit positions.
Let \(X_A \subseteq \{1,\dots,T\}\) be the set of 1-bit locations for document \(A\), and \(X_B\) for document \(B\).
From Step~1, both sets have size about \(s\): \(|X_A|\approx |X_B|\approx s\).
To estimate chance overlap for unrelated documents, we fix \(A\)'s bitmap and work with the set of 1-bit positions.
Since each bitmap has about \(s\) distinct 1-bits on average, we approximate an unrelated \(B\) as choosing \(s\) distinct bit positions uniformly at random without replacement from the \(T\) positions.
The overlap \(X = |X_A \cap X_B|\) counts how many of \(B\)'s chosen positions land in \(A\)'s \(s\) marked positions.
This problem maps to a hypergeometric distribution~\cite{siegrist_hypergeometric}, corresponding to drawing \(n\) items without replacement from a population of size \(N\) with \(K\) marked items; here \(N=T\), \(K=s\), and \(n=s\). 
Therefore,
\[
X \sim \mathrm{Hypergeom}(N{=}T,\;K{=}s,\;n{=}s).
\]

The expected overlap is
\[
\mathbb{E}[X]
= n\cdot \frac{K}{N}
= s\cdot \frac{s}{T}
= \frac{s^2}{T}.
\]

Intuitively, \(B\) makes \(s\) picks, and on any pick the chance of hitting one of \(A\)'s \(s\) one-bit locations is \(s/T\), so the expected number of hits is \(s\cdot (s/T)\). Equivalently, let \(I_j\) be the indicator that the \(j\)-th pick of \(B\) hits a marked position in \(A\). Then \(X=\sum_{j=1}^{s} I_j\), and by linearity of expectation,
\[
\mathbb{E}[X]
=
\sum_{j=1}^{s} \mathbb{E}[I_j]
=
\sum_{j=1}^{s} \Pr[I_j=1]
=
\sum_{j=1}^{s} \frac{s}{T}
=
\frac{s^2}{T}.
\]

With \(T=4096\) and \(s\approx 110.50\), this gives \(\mathbb{E}[X]\approx 3\): two unrelated documents share only about three 1-bits on average.
Bitmap Jaccard compares active bit positions by intersection-over-union:
\[
J_{\text{bitmap}}(A,B)
=
\frac{|X_A\cap X_B|}{|X_A\cup X_B|}
=
\frac{X}{|X_A|+|X_B|-X}
\approx
\frac{X}{2s-X}.
\]
Using \(X\approx 3\) yields a typical unrelated similarity around \(0.014\).

At our deduplication threshold \(J_{\text{bitmap}}\ge 0.7\), two documents of size \(\approx s\) would need a much larger overlap:
\[
\frac{X}{2s - X} \ge 0.7
\Longrightarrow
X \ge \frac{2\cdot 0.7}{1+0.7}\,s
\approx 0.8235\,s
\approx 91
\]
shared bits. Thus, a random non-duplicate pair would need \(X \ge 91\), even though \(\mathbb{E}[X]\approx 3\).

To make this concrete, we evaluate the hypergeometric tail using the integer approximation \(s\approx 110\).
Under \(X\sim \mathrm{Hypergeom}(4096,110,110)\), the exact tail probability is
\[
\Pr[X \ge 91]
=
\sum_{x=91}^{110}
\frac{\binom{110}{x}\binom{4096-110}{110-x}}{\binom{4096}{110}}
\approx 5.95\times 10^{-147}.
\]
Even over a large number of document pairs, the expected number of false positives due purely to bitmap collisions is effectively zero, which justifies using a 4{,}096-bit bitmap as a faithful surrogate signal at our operating thresholds.

\subsection{Hamming Distance vs. MinHash/Jaccard Agreement}
\label{appendix:hamming_minhash_example}

A concrete example makes this mismatch clear. Consider two documents with three hash values (shown here as integers for readability):

\ifdefstring{\papershortname}{FOLD}{
    \[
\left.
\begin{aligned}
  &\text{Doc 1: } \underbrace{23\;\;45\;\;67}_{\text{hash values}}\\[-0.25em]
  &\text{Doc 2: } \underbrace{22\;\;41\;\;12}_{\text{hash values}}
\end{aligned}
\right\}
\Rightarrow
\text{\#equal hash}=0,\;
\hat{J}=\frac{0}{6}=0\;
\]
    }{
     \[
\left.
\begin{aligned}
  &\text{D1: } \underbrace{23\;\;45\;\;67}_{\text{hash values}}\\[-0.25em]
  &\text{D2: } \underbrace{22\;\;41\;\;12}_{\text{hash values}}
\end{aligned}
\right\}
\Rightarrow
\text{\#equal hash}=0,\;
\hat{J}=\frac{0}{6}=0\;(\text{no similar})
\]
    }

Although no hash values match exactly and the MinHash/Jaccard agreement is hence 0, the Hamming distance tells a different story. The Hamming distance is computed on the bit strings of these hash values. Below we write the same integers in 8-bit binary (for illustration; a real implementation uses 32-bit hash values):

  \ifdefstring{\papershortname}{FOLD}{
           
\[
\left.
\begin{aligned}
&\text{Doc1: }\underbrace{\scriptsize 00010111\;\,00101101\;\,01000011}_{23\;\;45\;\;67\text{ in binary}}\\[-0.45em]
&\text{Doc2: }\underbrace{\scriptsize 00010110\;\,00101001\;\,00001100}_{22\;\;41\;\;12\text{ in binary}}
\end{aligned}
\right\}
=\underbrace{1 + 1 + 5}_{\text{bit flips}} = 7
\]
    }{
       \[
\left.
\begin{aligned}
&\text{D1: }\underbrace{\scriptsize 00010111\;\,00101101\;\,01000011}_{23\;\;45\;\;67\text{ in binary}}\\[-0.25em]
&\text{D2: }\underbrace{\scriptsize 00010110\;\,00101001\;\,00001100}_{22\;\;41\;\;12\text{ in binary}}
\end{aligned}
\right\}
\;\Rightarrow\;
d_H=\underbrace{1 + 1 + 5}_{\text{bit flips}} = 7
\]
    }

Across the three 8-bit values, the packed signature has $B=24$ bits, so the normalized Hamming similarity is
\[
1-\frac{d_H}{B}=1-\frac{7}{24}=0.708\approx0.71.
\]

Thus, the pair has $\hat{J}_{\text{MinHash}}=0$ due to zero exact hash matches, yet still exhibits $\approx 70\%$ similarity according to the normalized Hamming metric. This illustrates why Hamming distance is an unstable proxy for the Jaccard objective.

\end{document}